\documentclass[a4paper, 12pt]{article}
\usepackage{epsfig}
\usepackage{amssymb,amsmath,amsthm}
\usepackage{indentfirst}

\title{\textbf{Statistical detection of movement activities in a
human brain by separation of mixture distributions}}

\begin{document}
\date{}
\maketitle

\vspace{-2cm}

\begin{center}
{\large \bf  A.\,K.~Gorshenin\footnote{Institute of Informatics
Problems, Russian Academy of Sciences; Moscow State Institute of Radio Engineering, Electronics and Automation; Vavilova str., 44/2,
Moscow, Russia. Email: agorshenin@ipiran.ru},
V.\,Yu.~Korolev\footnote{Lomonosov Moscow State University;
Institute of Informatics Problems, Russian Academy of Sciences;
Leninskie Gory, Moscow, Russia. Email: victoryukorolev@yandex.ru},
A.\,Yu.~Korchagin\footnote{Lomonosov Moscow State University;
Leninskie Gory, Moscow, Russia. Email: sasha.korchagin@gmail.com},
T.\,V.~Zakharova\footnote{Lomonosov Moscow State University;
Leninskie Gory, Moscow, Russia. Email: tvzaharova@mail.ru},
A.\,I.~Zeifman\footnote{Vologda State University,
S.Orlova, 6, Vologda, Russia; Institute of Informatics Problems,
Russian Academy of Sciences; Institute of Socio-Economic Development
of Territories, Russian Academy of Sciences. Email: a\_zeifman@mail.ru} }
\end{center}

\bigskip

\small


\newcommand*{\E}{\mathbb E}
\newcommand*{\D}{\mathbb D}
\newcommand*{\R}{\mathbb R}
\renewcommand*{\Pr}{\mathbb P}

\small

{\bf Abstract\\} One of most popular experimental techniques
for investigation of brain activity is the so-called method of
evoked potentials: the subject repeatedly makes some movements (by
his/her finger) whereas brain activity and some auxiliary signals
are recorded for further analysis. The key problem is the detection
of points in the myogram which correspond to the beginning of the
movements. The more precisely the points are detected, the more
successfully the magnetoencephalogram is processed aiming at the
identification of sensors which are closest to the activity areas.

The paper proposes a statistical approach to this problem based on
mixtures models which uses a specially modified method of moving
separation of mixtures of probability distributions (MSM-method) to
detect the start points of the finger's movements. We demonstrate
the correctness of the new procedure and its advantages as compared
with the method based on the notion of the myogram window variance.

\medskip


\normalsize


\section*{Introduction}

The human brain is the center of the nervous system, the cerebral
cortex processes all information. Therefore, the research of brain
activity is one most important problem of modern medicine. In order
to investigate human mental activity and obtain new information
concerning the structure and relationships between the functional
areas in the brain, various statistical methods can be used.

A precise preoperative localization of irreplaceable areas in the
brain is the basis for planning of surgeries and minimization of
possible postoperative aftereffects of illness for patients with
brain disorders. In clinical practice, the localization of primary
motor cortex (M1) is one of most important and difficult problems,
especially the problem of detection of the hand activity's area in
M1. Due to various involvements of central nervous system (e. g., as
a result of epilepsy), the information concerning the configuration
of areas by the anatomical data is deficient.

Magnetoencephalography (MEG) is a method of preoperative
localization of various brain areas. It is often used in a
combination with magnetic resonance imaging (MRI) to localize
activity in the areas of the human brain. The method of evoked
potentials is one of most popular techniques to determine the exact
location of the motor areas of the cortex~\cite{McGillem1987,
Fabiani2007}.

The external actions and some movements of the subject can be
considered as the events for the analysis by the method of evoked
potentials. Such methodology leads to the increase of the
signal-to-noise ratio, and it allows to reveal the brain activity
corresponding to the events. The noise in this case is a
superposition of the physical noise (say, generated by the noises of
sensors, amplifiers, analog-to-digital conversion, external signals,
network interferences, etc.) and the physiological one (which is
just a background brain activity). The main problem of the method
relates to the detection of the starting point of a movement.

In brief, the scheme of the experiment is as follows. The subject
puts his hand on the table and taps by a forefinger for a few
minutes. His/her MEG signals and myogram are recorded during the
experiment. Additionally, the information about contacts with the
table surface can be registered (say, by the accelerometer,
photocell button, etc.). All signals are taken synchronously and
with a strictly fixed sampling frequency. MEG signals are dataset of
sensors located on the subject's head, each time series is called a
channel.

The main aim of the analysis of the experimental data is the
detection of the area of the cerebral cortex which is responsible
for the beginning of the movement. This is a particular case of the
so-called inverse problem of finding the source of a signal by the
characteristics of the field generated by the source. One of the
simplest solutions is to find the channel with the best response by
averaging parts of MEG signals over the starts of the movements.
Then, the corresponding curve could be chosen to improve the
signal-to-noise ratio. However, start points cannot be determined by
the MEG signals due to the part of noise in channel (it equals
$0.95$ and more). But the beginning of the movement can be found
from the myogram, and this is sufficient for the averaging of MEG.
At the same time, the values of a button can be used to adjust the
fact of movement in the neighborhood of some point.

In the paper \cite{Zakharova2012} an algorithm was proposed for
solving the problem under discussion. That algorithm was aimed at
the proper identification of reference points. For this purpose a
simple property of a myogram was used: its window variance
associated with muscle movements, due to the physiological
characteristics of the human muscles, exceeds significantly the one
related to the rest period (the window width is $30-50$ ms).
%

The simplest reasonable mathematical model for a myogram is a cyclic
non-stationary random process which can be represented as
\begin{eqnarray*}
\xi(t)=\sum_i ((s_i(t)+\varepsilon_i(t) + \theta_i(t))\mathbf{1}\{t_i\leqslant t<t_{i+1}\}),
\end{eqnarray*}
where the random process $s_i(t)$ corresponds to the signal
component related to the finger movement ($s_i(t) = 0$ out of the
movement period), $\varepsilon_i(t)$ is the rest noise (it equals
zero during a movement), $\theta_i(t)$ is the movement noise (it
equals zero during a rest period), in neurophysiology the
half-interval $[t_i, t_{i+1})$ is called an epoch (each epoch
includes the rest interval before the movement and  the movement
itself), $i$ is the epoch number.

The window variance obtained from the myogram is also a cyclic
non-stationary process, but much less contaminated with noise. The
transfer to the window variance allows to eliminate trends and to
emphasize rest-to-movement transition moments which are then
determined by the threshold processing. The method proposed in
\cite{Zakharova2012} demonstrated very high accuracy. However, due
to the noticeable non-normality of the distribution of noise, within
the method proposed in \cite{Zakharova2012} the thresholds were
determined rather artificially.

The present paper proposes some developments of the method of
\cite{Zakharova2012}. In the following sections we describe some
statistical procedures based on mixture models designed for precise
detection of the points corresponding to the beginning of movements.


\section*{Smoothing the signal by moving separation of finite mixtures}

To reveal the changes of the structure of the stochastic processes
in time, the so-called method of moving separation of mixtures (MSM
method) is successfully used. This method was proposed
in~\cite{Korolev2011}. As examples of efficient performance of this
method, the papers~\cite{Gorshenin2008, Gorshenin2013Alesund1,
Gorshenin2014} should be mentioned containing results for the
financial markets, for the traffic in information systems and for
the plasma turbulence correspondingly. The key point in this method
is that the volatility of the process is decomposed into two
components, dynamical and diffusive.

Within the framework of this method, the one-dimensional
distributions of the increments of the basic process are
approximated by finite location-scale mixtures of normal
distributions. The theoretical background of these models can be
found in \cite{Korolev2011}.

To analyze the dynamics of the changes in the stochastic process,
the problem of statistical estimation of unknown parameters of
distributions should be successively solved for a part of sample
which moves in a direction of astronomical time (i.e., the initial
sample is divided into sliding or moving sub-samples often called
windows). Typically, the window (sub-sample) size is fixed. Once the
analyzed parameters are obtained for a current location, the window
should be moved by one element of the initial sample (i.e., the
method will analyze the next sub-sample). This allows to detect all
possible changes in the behavior of components.

We suppose that the cumulative density function for corresponding
moment of time (location of a window) can be represented as
\begin{gather}
F(x)=\sum_{i=1}^k
\frac{p_i}{\sigma_i\sqrt{2\pi}}\int\limits_{-\infty}^{x}\exp\Big\{-\dfrac{(t-a_i)^2}
{2\sigma_i^2}\Big\}\,dt, \label{Mixture}\\
\intertext{where} \sum\limits_{i=1}^{k}p_{i}=1, \ \ \ p_{i}\geqslant
0.\label{Weights}
\end{gather}
(for all $x \in \R$,  $a_i\in\R$, $\sigma_i>0$, $i=1,\ldots,k$). The
model~\eqref{Mixture} is called a finite location-scale normal
mixture. The parameters  $p_{1},\ldots,p_{k}$ are weights
satisfying~\eqref{Weights}. The parameter $k$ is the number of
mixture components. The parameters $a_{1},\ldots,a_{k}$ are
associated with the dynamic component of the volatility (variance)
of the process, the parameters $\sigma_{1},\ldots,\sigma_{k}$ are
associated with the diffusion one, see~\cite{Korolev2011}. Namely,
if $Z$ is a random variable with distribution
function~\eqref{Mixture}, then its variance can be represented as
the sum of two components:
\begin{equation}\label{VolDec}
{\sf D}Z=\sum_{i=1}^{k}p_i(a_i-\overline
a)^2+\sum_{i=1}^{k}p_i\sigma_i^2,
\end{equation}
where
$$
\overline a=\sum_{i=1}^{k}p_ia_i.
$$
The first term on the right-hand side of~\eqref{VolDec} depends only
on the weights $p_i$ and the expected values $a_i$ of the components
of mixture~\eqref{Mixture}. Since $Z$ is an increment of the basic
process, then $a_i$ is the expected value of the increment, i. e.,
the trend component. Hence, the first component is the part of the
total variance (changeability) which is due to existing elementary
trends. It is called \emph{a dynamic} component of the variance. At
the same time, the second term on the right-hand side
of~\eqref{VolDec} depends only on the weights $p_i$ and the
variances $\sigma_i^2$ of components and represents the purely
stochastic \emph{diffusive} component of the total variance.

\section*{Detection of starting points from the myogram by MSM method within finite mixture model}

\begin{figure} [!h] 
\centerline{
\includegraphics[width=0.5\textwidth, height=0.4\textwidth]{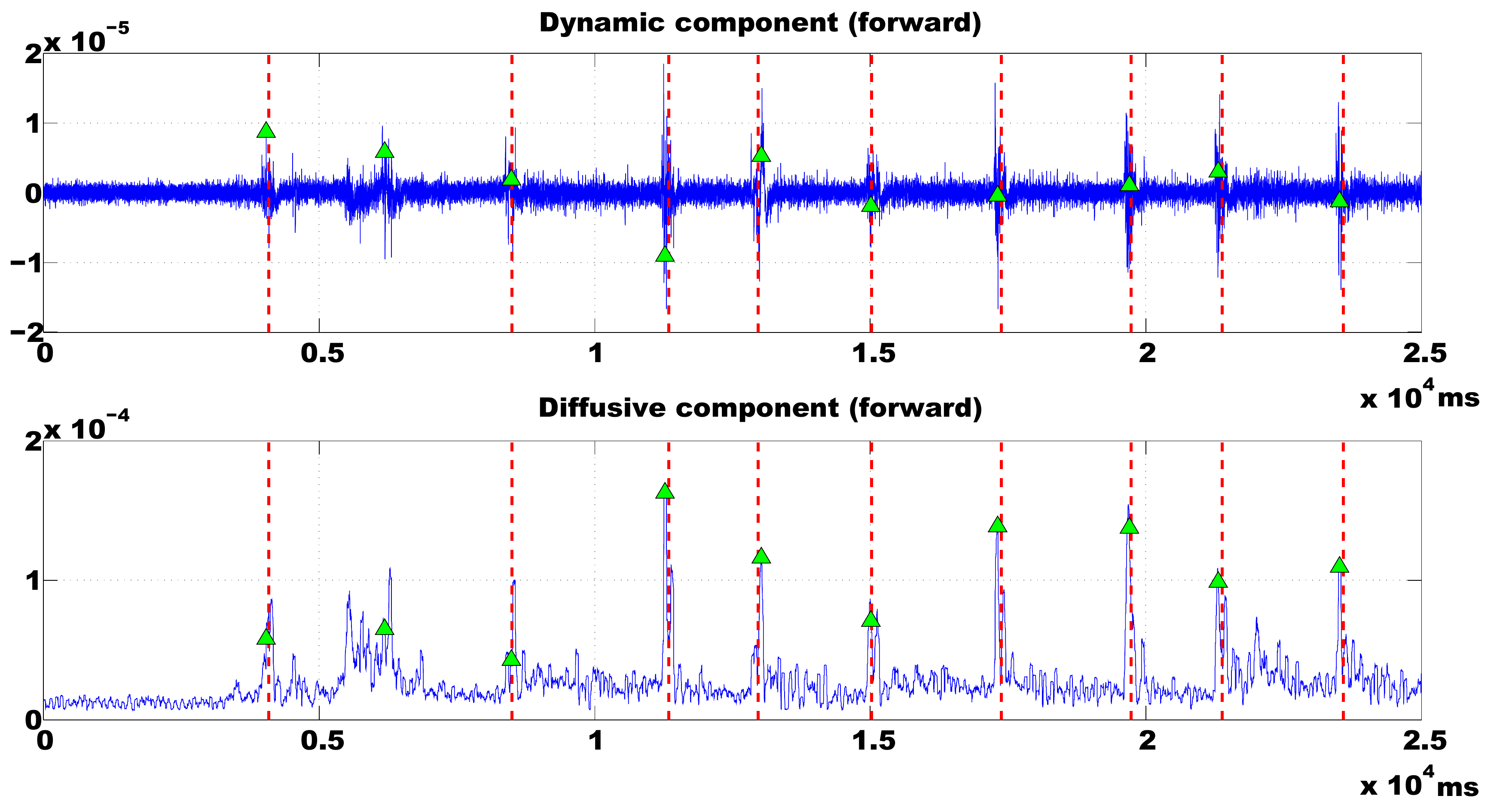}
\includegraphics[width=0.5\textwidth, height=0.4\textwidth]{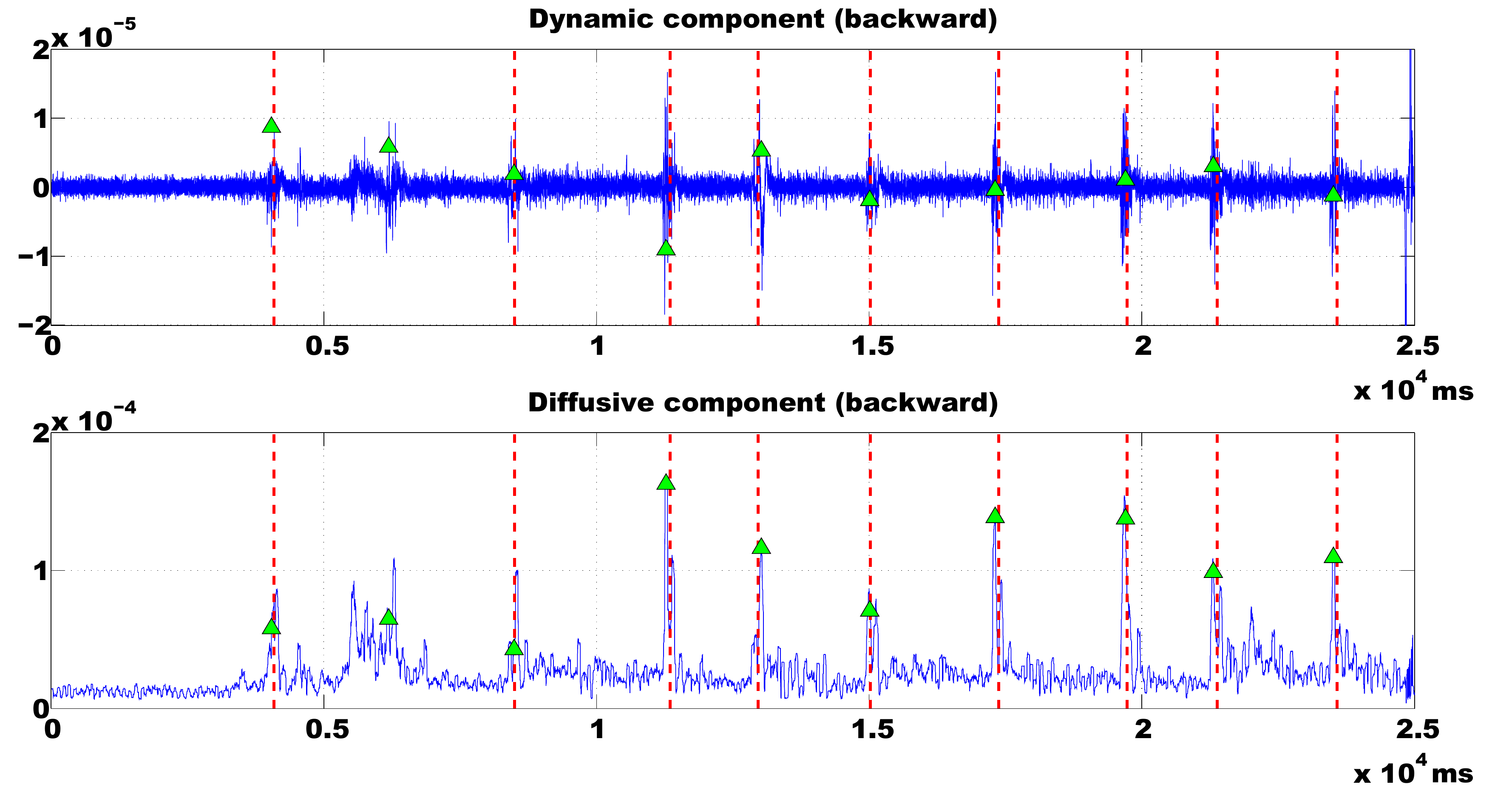}}
\caption{{\small The forward (left) and backward (right) direction
of analysis. The thin solid lines are components (a dynamic one is
located on top of the figure, a diffusive component is arranged on
below). The points of movements by method based on myogram window
variance and by method based on moving separation of mixtures: the
vertical red dashed lines and the green triangles
respectively.}}\label{Backward}
\end{figure}

\begin{figure*} [!h]
\begin{center}
\includegraphics[width=0.9\textwidth, height=0.4\textwidth]{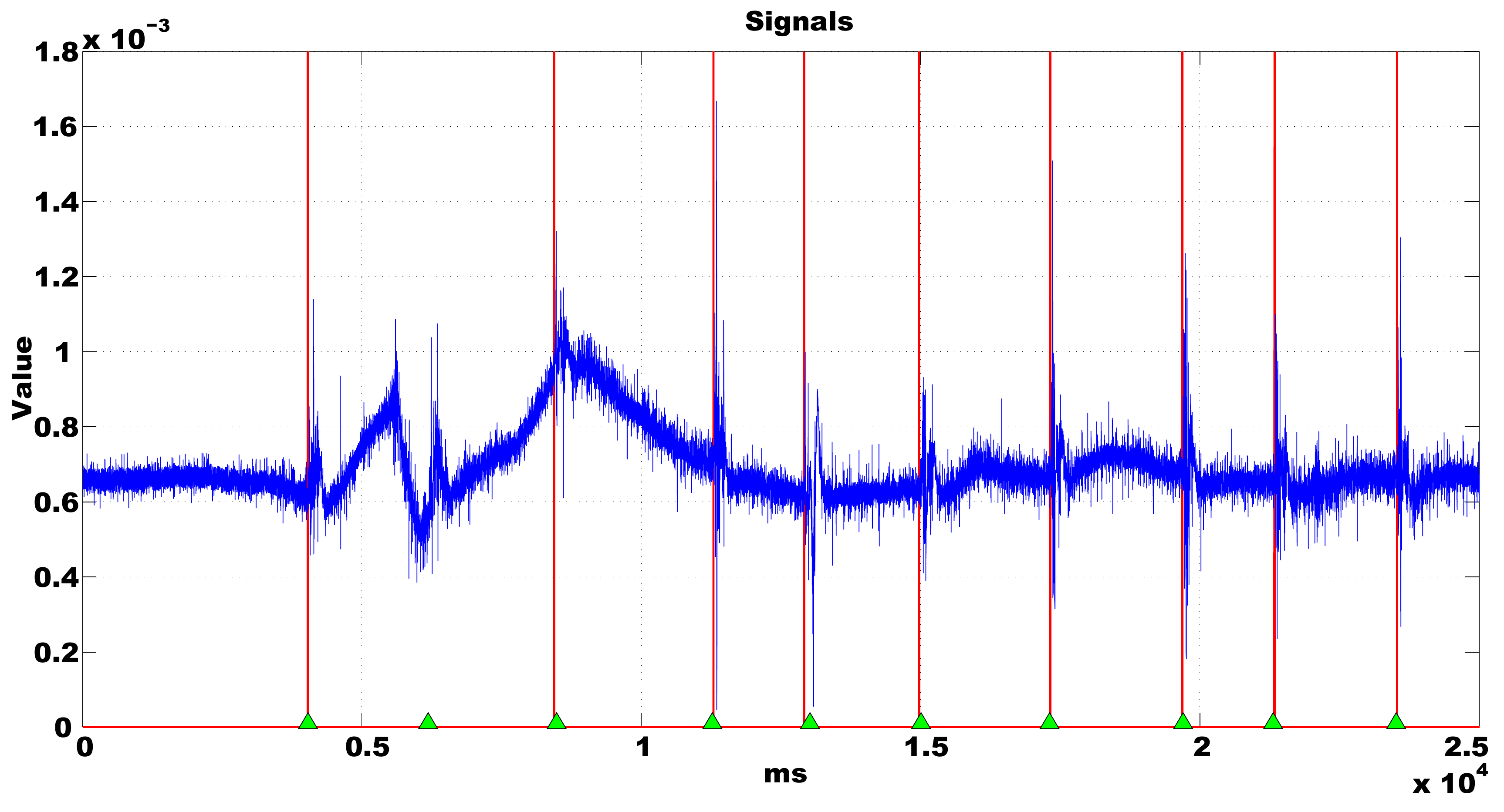}
\caption{{\small The myogram. The thin solid line demonstrates
signals of myogram, the vertical red solid lines and the green
triangles are the movement points for the window variance method
\cite{Zakharova2012} and moving separation of finite normal
mixtures, respectively.}}\label{AN_LH25}
\end{center}
\end{figure*}

The main idea of detection of movements by the MSM method is as
follows. First of all, the dynamic and diffusive components should
be estimated. In most situations, various modifications of usual
EM-algorithm are involved~\cite{Korolev2011}. The results of such a
decomposition are shown on fig.~\ref{AN_LH25} (the initial myogram
time series), and fig.~\ref{Backward} (the dynamic and diffusive
volatility (variance) components).

\begin{figure*} [!h]
\begin{center}
\includegraphics[width=0.8\textwidth, height=0.33\textwidth]{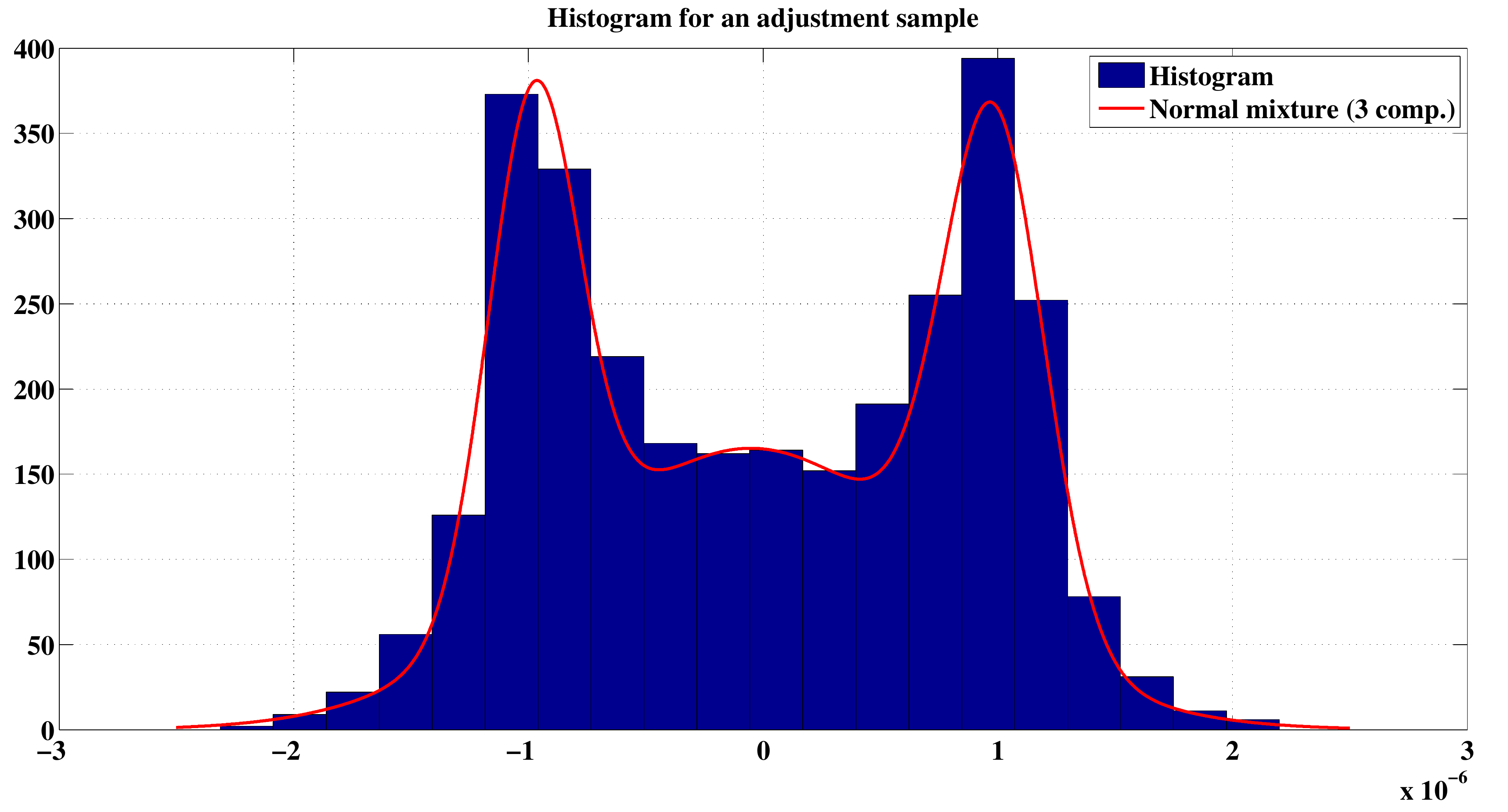}
\caption{Histogram of the values of the dynamic component ov
volatility for an adjustment sample (rest period) with the
approximating finite normal mixture.}\label{AdjustmentSample}
\end{center}
\end{figure*}

It seems more prospective to use the obtained dynamic component of
the total variance as the initial data (time series) for further
analysis, see below. It is very interesting that the distribution of
the values of the dynamic component obtained over the rest period is
very far from being normal, see fig.~\ref{AdjustmentSample}

One of key problems in detection of movements by the MSM method is a
possible time delay. The analysis is based on sub-samples, and the
influence of a new moment of time (one element of the sample) could
appear with a delay. But we are interested in the precise detection
of movements. The solution is based on idea of double sample
processing by MSM method: in forward and backward directions.
Comparing the probable points for both directions, we can find right
location of movements.

To detect the points of movements we use the myogram as the initial
data (sample). To avoid trends in the myogram (and in the dynamic
volatility component), the differences of successive elements of the
sample should be found. We tested various sizes of window (e.g.,
$20$, $30$, $50$) to compare the character of components. In the
paper we present the results for sub-samples of size equal to $50$
elements in each position of a window.

The algorithm described above was applied to the myogram. The
results for both directions are represented on Fig.~\ref{Backward}.
The $x$-axis corresponds to the time of experiments (in
milliseconds), the $y$-axis demonstrates the corresponding values of
components. The thin solid lines are components (above is the
dynamic component and below is the diffusive one).

We can see the balance statement of the subject at the beginning of
components approximately until the point $3000$. Note that it is
very important to use the diffusive component to determine the mode
due to possible ambiguities in the dynamic component. We could
exploit the balance statement of the subject to estimate the bounds
to be used for movements detection. Fig.~\ref{AdjustmentSample}
demonstrates histogram for the sample corresponding to the balance
statement of the subject. The distribution is a multimodal, it can
be approximated by the mixture model~\eqref{Mixture} too. So, the
standard technique based on the quantiles of unimodal distribution
(like a $w$-sigma rule for a normal distribution) cannot be used to
fix the bounds. Surely, the bounds can be chosen empirically with a
help of an additional information. For example, the data of
photocell button could promote to evaluate the bounds for all
signals in the experiment by a few movements in the beginning of a
myogram. In the following sections we discuss more valid statistical
approach based on the chi-squared tests.

The probable points of movements are determined by the crossings of
the bounds. Next we need to group the probable points of both
directions. It seems reasonable to classify groups with the help of
metric related with the window size. For example, all of probable
points which are located inside the interval with length
$j\times(\text{\rm window size})$, should be classified as a group.
And the starting point of a movement is the average of their time
locations.

Fig.~\ref{Backward} demonstrates the points of movements obtained by
the method based on myogram window variance and by the method based
on moving separation of mixtures: the vertical red dashed lines and
the green triangles, respectively. We can see a good compliance
between the results of the two statistical methods. The only
exception is the triangle near the mark $6200$~ms (the second
triangle from left). There are two reasons for that point. First, at
the beginning of experiment, the subject adjusts to the conditions
and can move not only his finger but his head or leg as well. But
this influences the myogram. Second, for the method based on myogram
window variance the information from photocell button was available.
It allows to omit some false moments. For the MSM method, we had no
data but the myogram.

We compared the results of the method for the components graphs. But
we analyzed differences of myogram. So, it needs to compare the
starting points of movements with a window
variance method for initial myogram as
is demonstrated in Fig.~\ref{AN_LH25}.

The $x$-axis corresponds to the time of experiments (in ms), the
$y$-axis demonstrates the corresponding values of the myogram. The
thin solid line demonstrates the signals of myogram, the vertical
red solid lines and the green triangles are the movement points for
two methods respectively. There are no changes for the conclusions
mentioned above.

It should be noted there are a few differences between forward and
backward results. The shape and values of backward diffusive
components are almost coincided with a forward one except the end of
graphs. For the backward mode it is an adjustment area for a
smoothed EM algorithm. A similar explanation is correct for the
dynamic components too. Also, a backward dynamic component is a
reflection of a forward one. It can be explained by the reason that
dynamic component is an expectation.

\section*{Detection of starting points by moving grid method from the dynamic component}

The main idea is to treat the dynamic component as a random process
(this is actually so) and process its distributions assuming that
they are normal mixtures themselves (see
fig.~\ref{AdjustmentSample}). To separate these mixtures we use the
modified two-step grid-based method~\cite{Korolev2014}, using only
its first step. As the result, we obtain 
the values of the weights $p_i$ of the $i$th node of the grid for
every window. We set the window size equal to $100$ and the window
shift equal to $1$.


In order to compare the vectors $\mathbf{p}_i$ of probabilities
$(p_{i,1},\ldots,p_{i,K})$ obtained on different nonintersecting
(for independence) windows, for each window with number $i > 100$ we
calculate the value
\begin{equation*}
z_i = \|\mathbf{p}_i - \mathbf{p}_{i-100} \| = \bigg[\sum_{j=1}^{K}
(p_{i,j}-p_{i-100,j})^2\bigg]^{1/2}.
\end{equation*}

Let us set the threshold value for $z_i$ in scope as $\theta = 0.97$
to highlight only extreme changes of the vector
$\mathbf{p}_i=(p_{i,1},...,p_{i,K})$ ($K$ is the number of nodes in
the grid). All $z_i > \theta$ are colored purple. Purple events are
grouped in tight groups so that every group is {\em continuous},
meaning that it consists of windows following one another. So we do
not have any issues with detecting groups instead of single events.

From each group we take the first value and shift it by $+150$
(window size used for this decomposition ($100$) + window size used
to obtain source data, the dynamic component ($50$)). We also take
into consideration that every group might have a {\em reflection}
after the detection value. We consider the next group as a
reflection if it is no further than $300$ windows apart from the
original group. Reflections are excluded from the detection process.

\begin{figure} [!h]
\centerline{
\includegraphics[width=0.48\textwidth, height=0.33\textwidth]{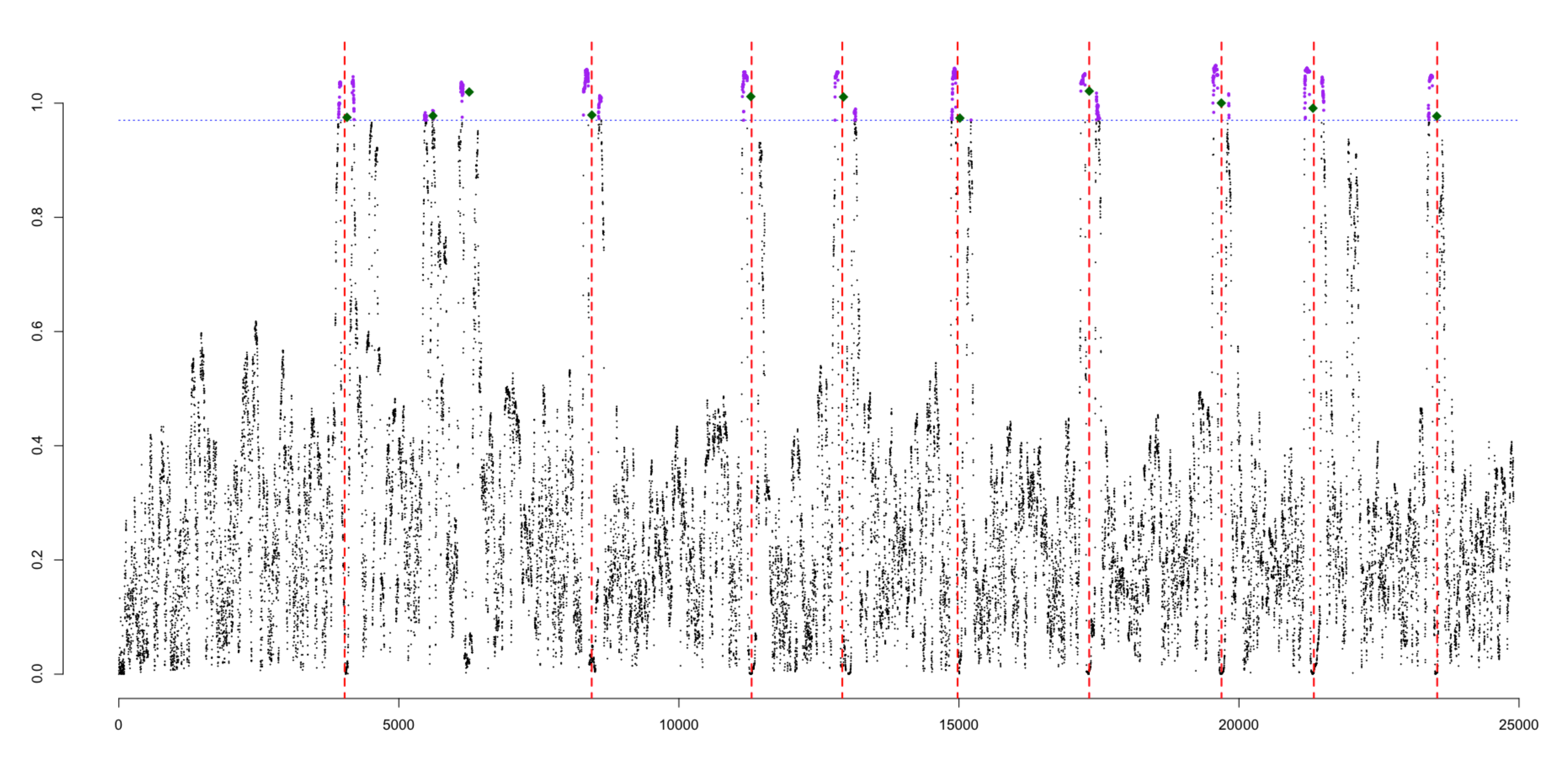}
\includegraphics[width=0.48\textwidth, height=0.33\textwidth]{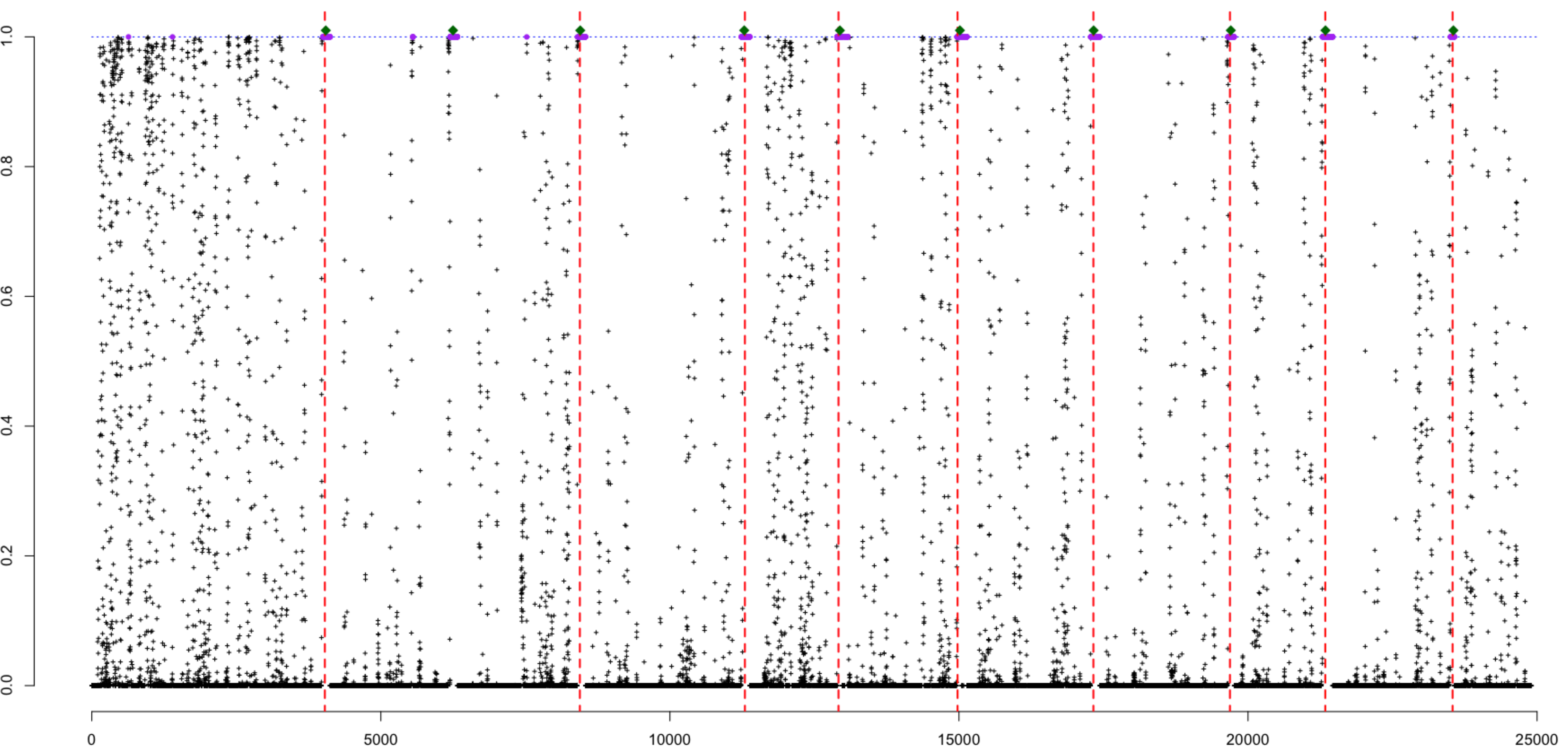}}
\caption{\small Movement detection using the myogram volatility
dynamic component decomposition, $z_i$-test (left) and chi-square
test (right)}\label{pV}
\end{figure}

Simple algorithm above produces the following time points (marked
green, see Fig.~\ref{pV}) (left): $4074$,  $5608$,  $6256$,  $8446$,
$11284$, $12938$, $15017$, $17327$, $19685$, $21321$, $23531$. The
actual events are (red lines): $4032$,  $8443$, $11298$, $12917$,
$14976$, $17326$, $19688$, $21337$, $23539$.

As we can see, between the first real event and the second real
event we detect a couple of auxiliary events. All other points are
estimated with the average accuracy of 7 ms, which is a great
result.

The previous experiment led us to a conclusion that the vector
$\mathbf{p}$ of weights is usually volatile. Let us focus on the
events themselves. If we analyze the behavior of vector $\mathbf{p}$
{\em inside} the detected events, we notice that it is changing
slower than it does {\em outside} of detected events. As the benefit
from this fact we use the chi-square test to detect the periods of
``stability'', which are much easier to determine.

For each window, we calculate the $P$-value of chi-square test with
5 bins. Once this is done, we will select the windows where p-value
is almost equal to $1$, i.e. greater than $0.9999$. These windows
are marked purple, see Fig.~\ref{pV} (right).


We are interested in long periods of stability, not random (noisy)
events, so we filter the groups that are less than 50 milliseconds.
Once we apply this filtering, we take the first point in each group.
In the previous experiment we added a fixed value of $+150 $ to each
event. In this case the stability period is detected only once the
window is {\em fully over} the event, so we need to add only $+50$
milliseconds (the window size used to obtain source data, the
dynamic component).

Using the method above, the following detected points are produced
(marked green, shifted above purple intervals to make them more
visible): $4048$,  $6249$, $8450$, $11286$, $12941$, $15017$,
$17330$, $19703$, $21340$, $23553$. The actual events are (red
lines): $4032$, $8443$, $11298$, $12917$, $14976$, $17326$, $19688$,
$21337$, $23539$.

As before, the method detects events between the first actual event
and the second one, but in this case we have only one false
detection, which means that this method is better. The average
accuracy of real events estimation is $12$~ms.

\section*{Detection of starting points by moving grid method from the myogram}

We applied the two-step decomposition algorithm \cite{Korolev2014}
to the source data, the myogram time series, directly. We used two
different distribution families: Generalized Hyperbolic (GH-)
\cite{Korolev2013} and Generalized Variance Gamma (GVG-)
distributions \cite{KorolevZaks2013} to fit the data. The cumulative
distribution functions of both of these laws are special normal
variance-mean mixtures of the form
$$
F(x)=\int\limits_{0}^{\infty}\Phi\Big(\frac{x-\alpha
u}{\sigma\sqrt{u}}\Big)dG(u), \ \ \ x\in\mathbb{R},
$$
where $\Phi(x)$ is the standard normal distribution function,
$\Phi(x)=\frac{1}{\sqrt{2\pi}}\int_{-\infty}^xe^{-t^2/2}dt$ and if
$F$ is the GH-distribution function, then $G$ is the generalized
inverse Gaussian distribution function \cite{J} whereas if $F$ is
the GVG-distribution function, then $G$ is the generalized gamma
distribution function \cite{Stacy1962}.

Both distribution families demonstrated good fit, but the
GVG-distributions have slightly better $P$-values when applying the
chi-square test. Figures below demonstrate the comparison of fitting
GVG-distributions and GH-distributions for two randomly chosen
windows. Based on the above, for further analysis we used only the
GVG-distributions.

\begin{figure} [!h]
\centerline{
        \includegraphics[width=0.45\textwidth]{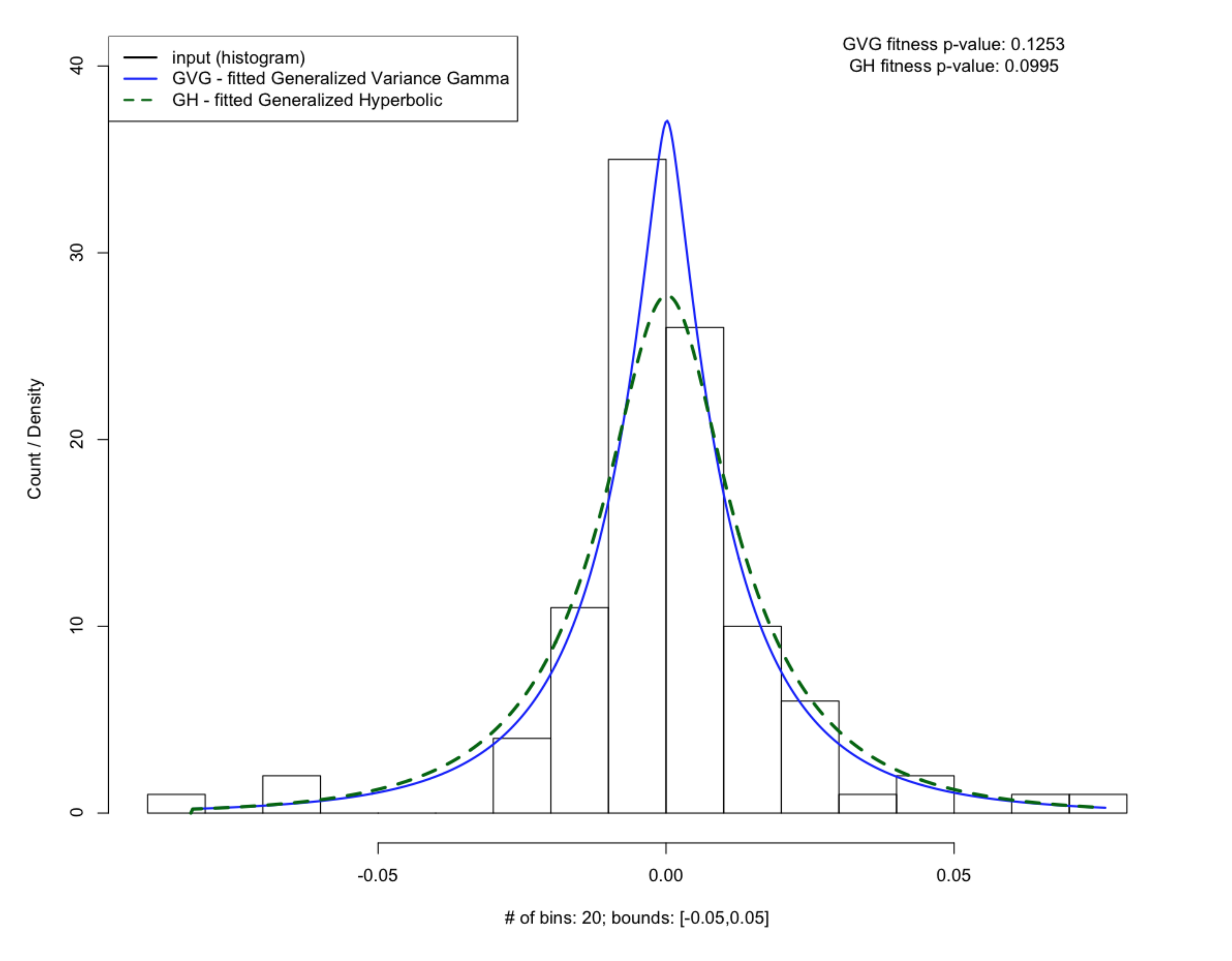}
        \includegraphics[width=0.45\textwidth]{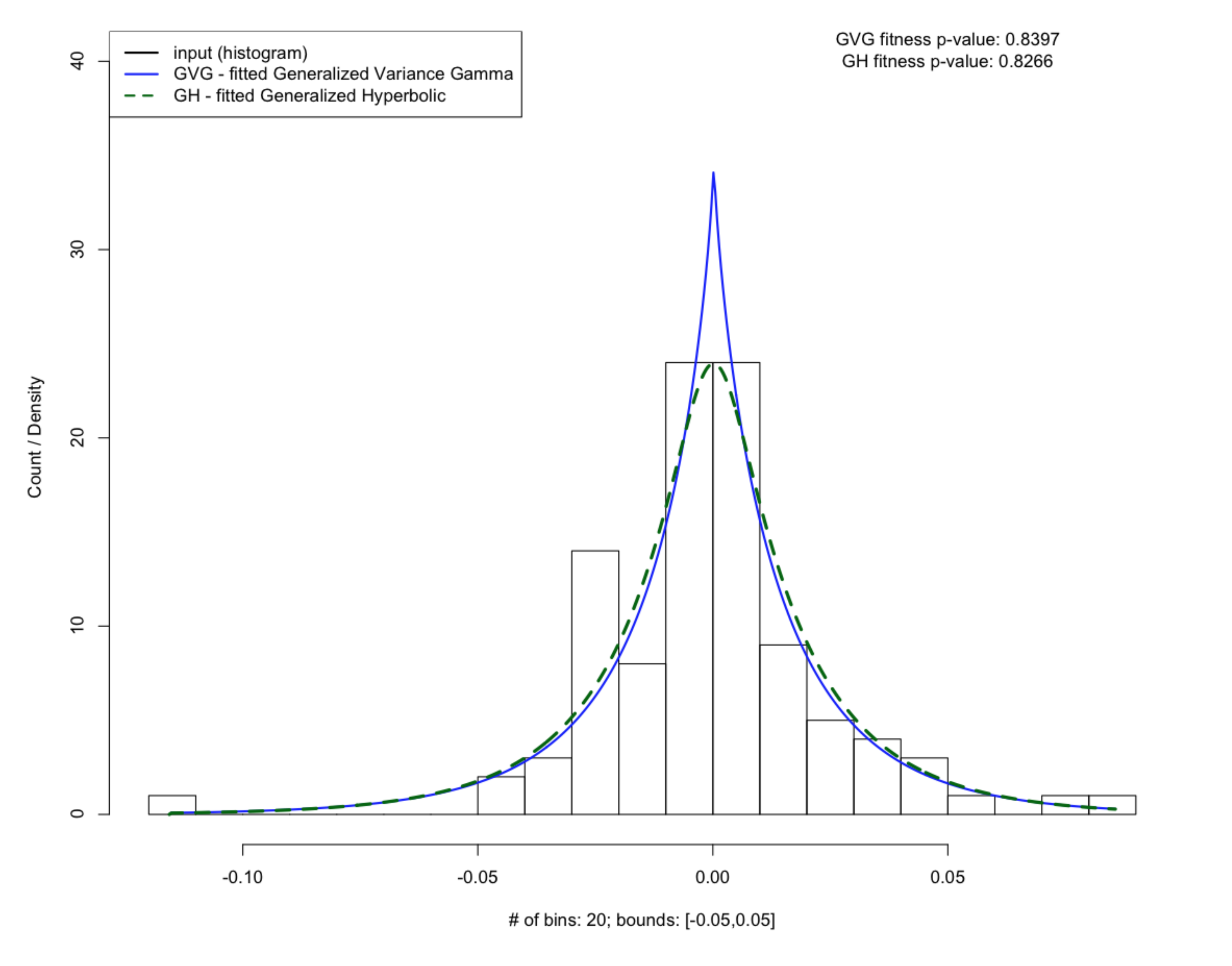}}
        \caption{\small GH vs GVG goodness-of-fit comparison, window no. 19251 (left) and window no. 10951 (right)}\label{19251}
  \end{figure}

We have a particular interest in the $\alpha$ parameter. It is
volatile, but near the events it tends to rapidly decrease, and then
to rapidly grow, {\em consistently} demonstrating large absolute
values. This means that we are able to detect events by watching the
average absolute values of $\alpha$.

\begin{figure} [!h]
    \begin{center}
        \includegraphics[width=0.9\textwidth, height=0.33\textwidth]{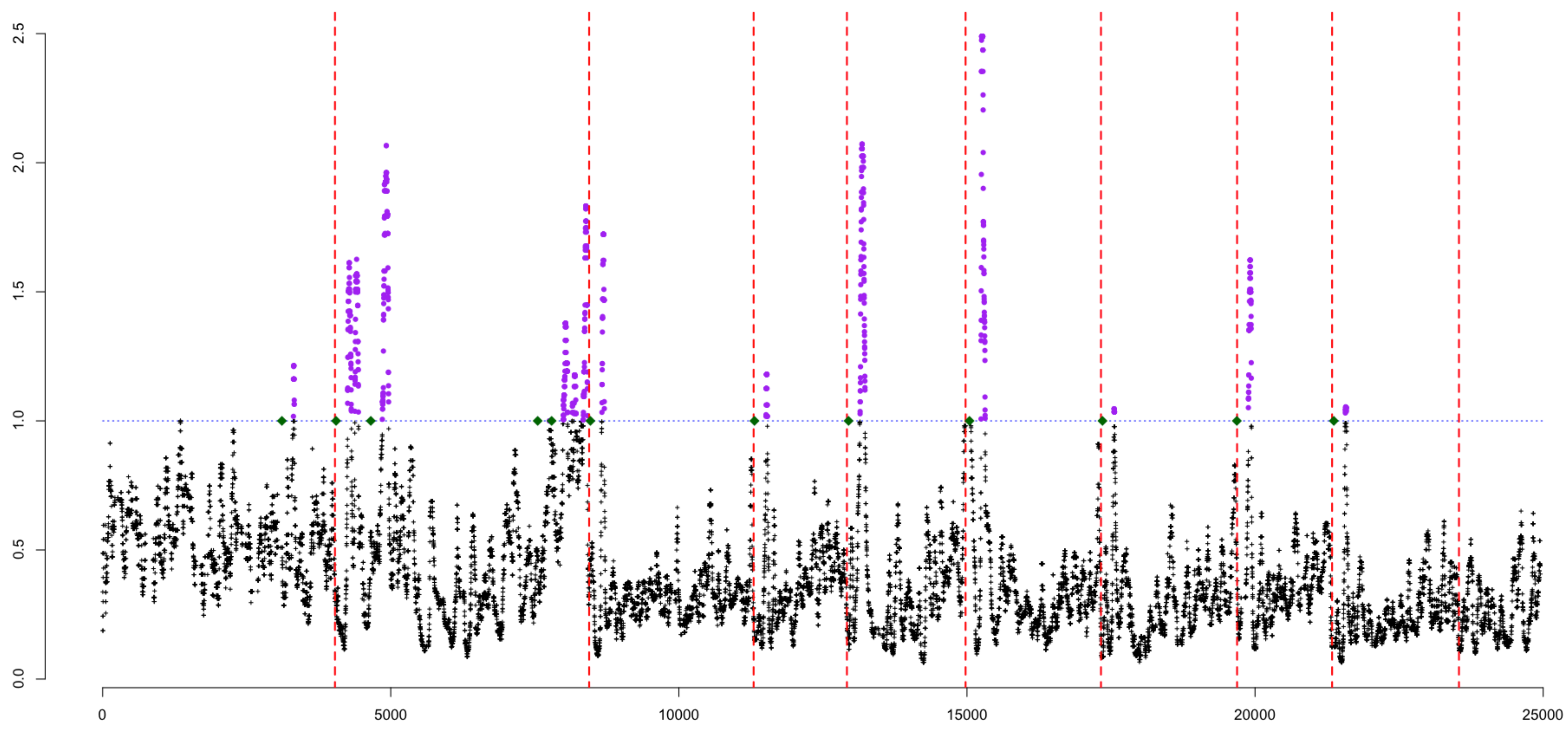}
        \caption{\small Movement detection using the $\alpha$ parameter of the GVG-distribution} \label{alpha}
    \end{center}
\end{figure}

We use threshold $1.0$ to select only high values (highlighted in
purple, see Fig.~\ref{alpha}). Similar to the cases above, we will
group the data and select only the first value in a group. To
produce final prediction, we need to subtract $200$ from the values
($100$, the initial windows size + $100$ for average $\alpha$
calculation).

The result is: $3111$,  $4051$, $4654$, $7550$, $7792$, $8465$,
$11312$, $12945$, $15044$, $17352$, $19684$, $21367$. The actual
events are (red lines): $4032$,  $8443$, $11298$, $12917$, $14976$,
$17326$, $19688$, $21337$, $23539$.

As we can see, the ``$\alpha$-method'' detected false events around
first and second actual events. Also, the detection of last event is
missing. Other events are estimated very accurately ($\sim 25$~ms).

Other metrics calculated for particular GVG-distributions can also
be used. Any numerical characteristics (moments, quantiles,
assymetry, curtosis, etc.) can be calculated and used for detection,
if they demonstrate specific behavior before/after or within the
analyzed events.

\section*{Conclusions}

The paper demonstrates the efficiency of the proposed statistical
methods for solving important medical problem. This method is based
on mixture models and implements numerical procedures of separation
of mixtures. Different techniques realizing this method are
discussed. For example, the method based on MSM approach could
involve the processing of additional data (from accelerometer and
photocell button) to precise location of its points. In addition,
using model of probability mixture, we can obtain a convenient tool
for further theoretical researches in the important field of the
modern medicine. The methods could be used for processing various
types of signals of human brain.

\section*{Acknowledgements}

The research is supported by the Russian Science Foundation, grant
14-11-00364.

\end{document}